\documentclass[useAMS,usenatbib]{mn2e}
\usepackage{amssymb,amsmath} 
\usepackage[dvips]{graphicx}

\newcommand{\msun}{\mbox{$M_{\sun}$}}

\newcommand{\kep}{\mbox{\textit{Kepler}}}

\title[Validation of Kepler-21b using PAVO/CHARA]
{Validation of the Exoplanet Kepler-21b using PAVO/CHARA Long-Baseline Interferometry}

\author[D. Huber et al.]{Daniel Huber$^{1,2}$\thanks{NASA Postdoctoral Program Fellow, daniel.huber@nasa.gov}, 
Michael J. Ireland$^{1,3,4}$, Timothy R. Bedding$^{1}$, Steve B. Howell$^{2}$, \and Vicente Maestro$^{1}$, 
Antoine M\'erand$^{5}$, Peter G. Tuthill$^{1}$, Timothy R. White$^{1}$, \and
Christopher D. Farrington$^{6}$, P. J. Goldfinger$^{6}$, Harold A. McAlister$^{6}$,
\and Gail H. Schaefer$^{6}$, Judit Sturmann$^{6}$, Laszlo Sturmann$^{6}$, \and Theo A. ten Brummelaar$^{6}$, 
and Nils H. Turner$^{6}$ \\
$^{1}$Sydney Institute for Astronomy (SIfA), School of Physics, University of Sydney, NSW 2006, Australia\\
$^{2}$NASA Ames Research Center, Moffett Field, CA 94035, USA \\
$^{3}$Department of Physics and Astronomy, Macquarie University, NSW 2109, Australia \\
$^{4}$Australian Astronomical Observatory, PO Box 296, Epping, NSW 1710, Australia \\
$^{5}$European Southern Observatory, Alonso de Cordova 3107, Casilla 19001, Vitacura, Santiago 19, Chile \\
$^{6}$Center for High Angular Resolution Astronomy, Georgia State University, PO Box 3969, Atlanta, 
GA 30302, USA \\}

\begin{document}

\date{Accepted --. Received --; in original form --}


\maketitle

\label{firstpage}

\begin{abstract}
We present long-baseline interferometry of the \kep\ exoplanet host star HD\,179070 (Kepler-21) using 
the PAVO beam combiner at the CHARA Array. The visibility data are consistent with a 
single star and exclude stellar companions at separations $\sim$\,1--1000\,mas ($\sim$\,0.1--113\,AU) and
contrasts $<$\,3.5 magnitudes. This result supports the 
validation of the 1.6\,$R_{\earth}$ exoplanet Kepler-21b by \citet{howell12} and 
complements the constraints set by adaptive optics imaging, speckle 
interferometry, and radial velocity observations to rule out false-positives due to 
stellar companions. We conclude that long-baseline interferometry has strong potential 
to validate transiting extrasolar planets, particularly 
for future projects aimed at brighter stars and for host stars where radial velocity 
follow-up is not available.
\end{abstract}


\section{Introduction}
The NASA \textit{Kepler} Mission aims to find extrasolar planets in the habitable zones of solar-type 
stars through the detection of brightness dips as planets cross the stellar disc. 
While Kepler has been highly successful in finding exoplanet candidates, 
ground-based follow-up observations are important to confirm the detections. 
The most common astrophysical false positives for Kepler involve stellar companions 
that remain unresolved due to Kepler's large pixel size ($\sim$\,4''). False positives 
can be divided into companions that are physically 
bound to the target star (hierarchical triple systems) and companions that are 
either in the foreground or the background of the target due to chance alignment 
(blends). In both cases, a transit-like shape can be mimicked by eclipses of a stellar 
companion or transits of a planet around the secondary companion.

For large (Jupiter- and Neptune-sized) planets, candidates can often be confirmed 
using radial velocity observations, giving a direct estimate of the planet's mass 
\citep[see, e.g.,][]{borucki10,koch10,latham10,cochran11}, while transit timing variations can be 
used to confirm planets in multiple systems \citep[see, e.g.,][]{fabrycky12,ford12,steffen12}.
For many super-Earths and Earth-sized planets, however, the Doppler signature is typically too small 
compared to the intrinsic stellar variability, and transit timing variations might not be detected. 
In these cases, candidates are validated by excluding as many  
false-positive scenarios as possible. The first stage in this process
uses the Kepler data to detect signatures of stellar companions through
photocenter shifts and the comparison of 
transit depths \citep[see, e.g.,][]{batalha10b,bryson10}, followed by statistical 
modeling of potential blending scenarios \citep[see, e.g.,][]{torres11,fressin11,morton11}. 
These constraints are then combined with ground-based follow-up observations such as spectroscopy, 
speckle interferometry and adaptive optics imaging \citep[see, e.g.,][]{gautier10,howell11}.
High-angular-resolution observations using long-baseline interferometry offer a 
powerful tool to complement these methods and extend the parameter range that can be excluded, 
particularly for close-in (both bound and unbound) companions.

\citet{howell12} recently reported the detection of Kepler-21b, a 1.6\,$R_{\earth}$ planet 
in a 2.8\,d orbit around the bright ($V=8.3$) F6IV star HD\,179070. 
Extensive follow-up observations have been used to rule out a false positive detection 
in this system. Speckle interferometry rules out any unbound or bound stellar companions 
at separations $>$\,0.2'' ($\gtrsim\,22.6\,$AU) for contrasts up to 5\,mag, and at separations 
$>$\,0.05'' ($\gtrsim\,5.6\,$AU) for contrasts up to 4\,mag. 
Adaptive optics imaging revealed a $\sim$14.5\,mag companion at a separation of 0.7'' which, 
however, was found unable to mimick the observed transit shape due to its low mass.
Radial-velocity time series were also obtained, and showed 
no significant variation over 85\,days at a 5.6\,m/s level. 
Putting all these constraints together, the only possibilities escaping direct detection are 
close-in bound companions within $\sim$\,4\,mag of HD\,179070 in nearly face-on orbits (causing no 
detectable RV signature) with a similar radial velocity to the target star, and close-in blends within 
$\sim$\,7\,mag that may have escaped spectroscopic detection.
Here, we present long-baseline interferometry observations to complement and extend the validation 
efforts by \citet{howell12}.

\section{Observations}
We have observed HD\,179070 as part of our interferometric follow-up campaign of \kep\ stars 
using the PAVO (Precision Astronomical Visible Observations) beam combiner \citep{ireland08} at the 
CHARA (Center for High Angular Resolution Astronomy) Array \citep{brummelaar04}. 
PAVO is a three-beam pupil-plane beam combiner optimized for high sensitivity and angular resolution, 
recording visibilities over a spectral bandpass 
of $\sim\,$650--800\,nm with a limiting magnitude in typical seeing conditions of $R\lesssim8$\,mag. 
Using baselines reaching up to 
330\,m, PAVO/CHARA is 
capable of resolving angular sizes down to $\sim 0.3$\,mas.
For more details on the instrument and data reduction, we refer to \citet{ireland08}.

\begin{table}
\begin{small}
\begin{center}
\caption{Spectral types, photometry and expected angular sizes of stars in this study. Brackets 
indicate the last two digits of the $1\sigma$ uncertainties.}
\vspace{0.1cm}
\begin{tabular}{l c c c c c c}        
\hline         
HD  	& Sp.T.		& $V$ 		& $K$ & $\theta_{V-K}$ (mas) \\	
\hline
179070  		& F6IV  &  $8.262(11)$  	&   $6.945(18)$  &    $0.169$(09)  		\\
\hline
174260  		& B8V 	&  $7.323(05)$  	&   $7.465(18)$  &    $0.103$(05) 		\\
178591$^{1}$  	& B5V 	&  $7.130(57)$  	&   $7.191(24)$  &    $0.119$(06)		\\
183204  		& A0V 	&  $7.425(07)$  	&   $7.386(21)$  &    $0.110$(06)  		\\
\hline
\end{tabular} 
\label{tab:cal} 
\flushleft $^{1}$ HD\,178591 is an ellipsoidal variable with an amplitude of 
$\sim\,0.05$\,mag.
\end{center}
\end{small}
\end{table}

HD\,179070 was observed on 2 July 2011 in two-telescope mode using the S1-W1 (278\,m) baseline 
in excellent seeing conditions. Two scans were obtained, which were interleaved with observations of 
three different calibrator stars. Using various catalogs available in the literature,  
calibrators are typically chosen to be single field stars in close vicinity ($<10^{\circ}$) 
to the target star. 
Table 1 lists the spectral types, photometric properties, and expected sizes of all stars 
in this study. The predicted sizes were calculated using the $(V-K)$ relation given by \citet{kervella04}.
$V$ magnitudes have been extracted from the Tycho catalog \citep{perryman97} and 
transformed into the Johnson system 
using the calibration of \citet{bessell00}. $K$ magnitudes have been obtained from the 2MASS 
catalog \citep{cutri03,skrutskie06}. 
We have tested the photometry for reddening by comparing the observed $(B-V)$ colors with a 
list of intrinsic colors as a function of spectral type given by \citet{schmidtkaler82}. 
The observed colors have been found to be compatible with the spectral types for all stars except for 
the calibrator HD\,178591, which is classified as an ellipsoidal variable by Hipparcos. 
We have accounted for this by adding a systematic 
uncertainty corresponding to the observed variability amplitude ($\sim\,0.05$\,mag) 
to the statistical photometric uncertainty in the $V$ band. 
Note that the potential companion of HD\,178591 can be 
expected to have a negligible influence on the interferometric measurements. 
The final uncertainties for the 
predicted angular diameters were calculated by adding a conservative 5\% calibration uncertainty 
in quadrature to the photometric uncertainty.

\section{Results and Discussion}

\begin{figure}
\begin{center}
\resizebox{\hsize}{!}{\includegraphics{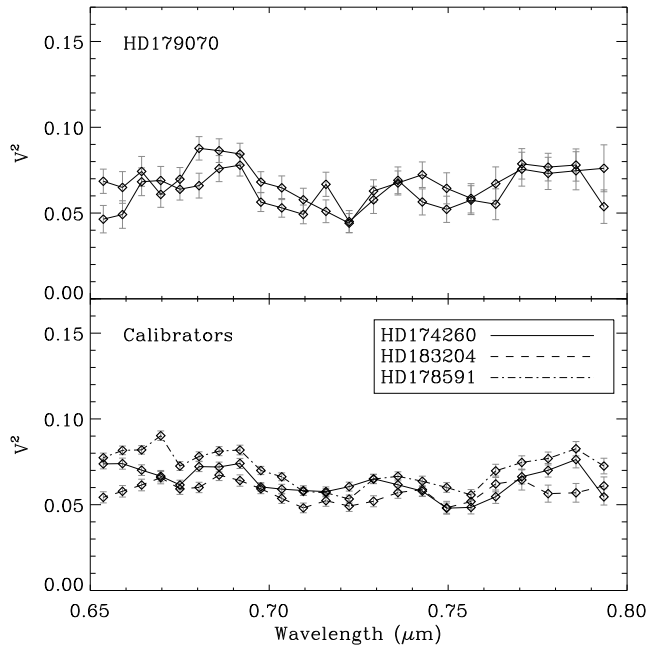}}
\caption{Raw visibility versus wavelength for the target HD\,179070 (top panel) and the 
calibrator stars (bottom panel).}
\label{fig1}
\end{center}
\end{figure}

\begin{figure*}
\begin{center}
\resizebox{\hsize}{!}{\includegraphics{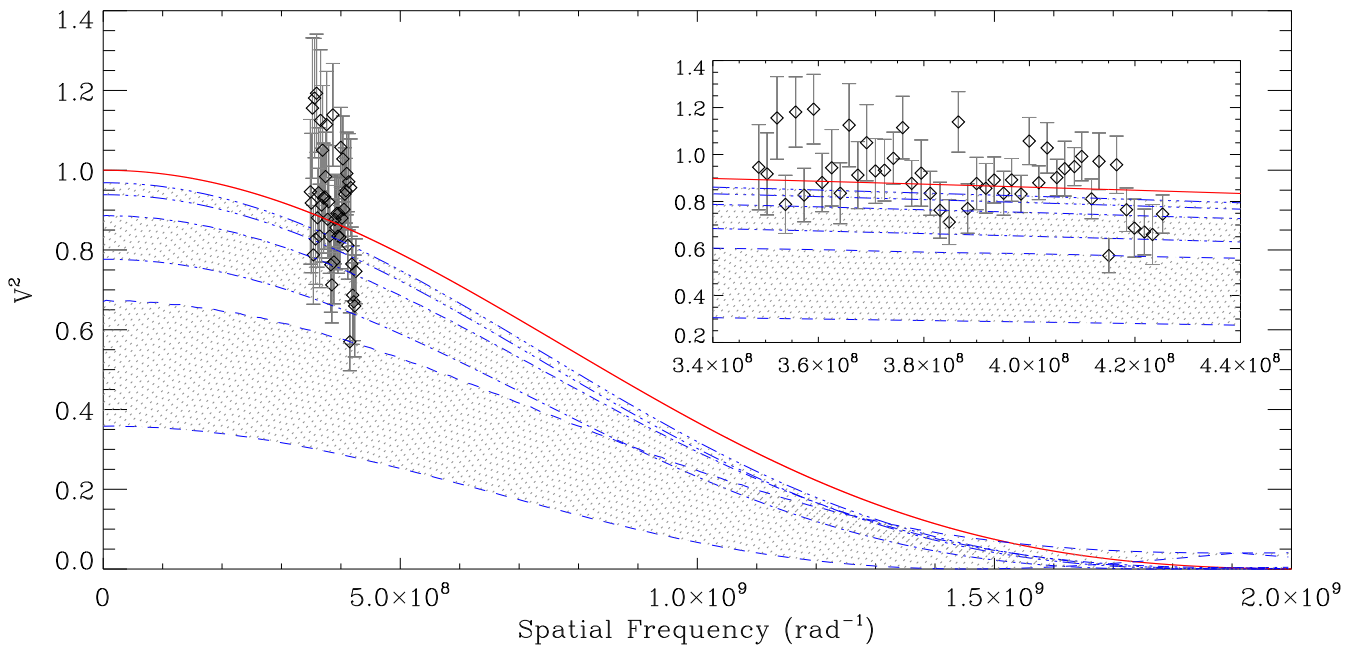}}
\caption{Calibrated squared visibility versus spatial frequency for HD\,179070. The red solid line 
shows the best fitting single disc model with a diameter of $\theta_{\rm LD}=0.13\pm0.02$\,mas. 
The areas marked by blue dashed, dashed-dotted and dashed-triple-dotted lines show the 
range of minimum squared visibilities expected for companions $>1$\,mas with contrasts of 1.5, 3 
and 4.5 magnitudes, respectively. The inset shows a close-up of the data.}
\label{fig2}
\end{center}
\end{figure*}

Figure 1 shows the raw squared visibility 
measurements across the PAVO passband for the target and the calibrators. As expected from their 
predicted sizes, the visibility levels of the calibrators are very similar. Note that the raw 
measurements are considerably lower than the expected visibilities ($\sim 0.9$) 
due to atmospheric turbulence and optical aberrations. 
HD\,179070 remains practically unresolved 
in our observations, with raw visibilities at very similar levels to the 
calibrator stars, and no significant visibility change as a function of wavelength. 
More importantly, 
the similarity between target and calibrator scans shows that the visibility curve of HD\,179070 is 
consistent with a single star. 
For any companion for which the interference patterns (fringe packets) of the 
two stars overlap in delay space, a periodic visibility modulation 
would be observed, while any incoherent flux from companions at larger separations 
would cause a drop in the observed visibility \citep[see, e.g.,][]{monnier03,brummelaar07}. 
The interferometric field of view for each case depends on the coherence length and 
the projected baseline \citep{brummelaar95}, and for our observations 
corresponds to $<$30\,mas and 30--1000\,mas, respectively.

To illustrate this more clearly, Figure 2 shows the calibrated squared visibility data of HD\,179070 
as a function of spatial frequency. Visibilities were calibrated by dividing the calibrator data 
by the predicted sizes to obtain a system visibility, which was then used to correct 
the target visibility. We then fitted the limb-darkened disc model given by \citet{hanbury74b} 
to the data using the method described by \citet{derekas11}. We used 
a linear limb-darkening coefficient for the $R$-band of $\mu_{R}=0.5197$, derived from the 
closest matching grid point of \citet{claret11} to the stellar parameters presented by 
\citet{howell12}.
The red line in Figure 2 shows the best-fitting single-disc model, yielding a 
diameter of $\theta_{\rm LD}=0.13\pm0.02$\,mas with a reduced $\chi^2=1.4$. 
This diameter agrees with the diameter of $0.15\pm0.01$\,mas constrained from the asteroseismic radius 
\citep[$R/R_{\sun}=1.86\pm0.02$,][]{howell12} 
and Hipparcos parallax \citep[$\pi=8.86\pm0.58$\,mas,][]{leeuwen07}. 
The areas marked by blue dashed, dashed-dotted, and dashed-triple-dotted lines illustrate 
the minimum squared visibilities expected for a stellar companion with a contrast of 
1.5, 3 and 4.5 magnitudes compared to HD\,179070. Deep minima correspond to 
close-in companions (1--30\,mas) which 
will show a periodic variation across the PAVO passband, while shallower minima correspond to 
wide companions ($>30$\,mas), for which the variation is unresolved within the spectral 
resolution of PAVO and hence an overall drop in visibility would be observed.
Figure 2 suggests that the PAVO observations rule out any 
stellar companions at contrasts of $\lesssim$\,3 magnitudes. 
Note that for stellar companions at even closer separations ($\lesssim 0.1$\,AU), the PAVO data 
would solely exclude companions along the baseline 
vector at the time of observation, which in this case spans only one epoch.
We also note that the faint $\sim 14.5$\,mag companion at $\sim$\,0.7\,''
detected by \citet{howell12} has negligible influence on our measurements.

To establish the magnitude limit more quantitatively, we performed $10^{5}$ simulations as follows. 
For each iteration, we used the spatial frequencies of our 
observations to generate a synthetic binary model consisting of a primary with the expected 
size of HD\,179070 and an unresolved secondary, with a separation and contrast drawn 
from uniform distributions between 1--50\,mas and 1--6\,mag, respectively. We then added white 
noise to each data point with a standard deviation corresponding to our estimated relative measurement 
uncertainties. For each simulated dataset, we then fitted a binary model to the data and compared 
the $\chi^{2}$ value to the one calculated from a single disc model with the expected size of 
HD\,179070. The simulations showed that in $>99$\% of all cases, the binary model yielded a 
significantly ($>\,3\,\sigma$) better fit for contrasts below 3.6 magnitudes. This 
limit has been found to be only weakly dependent on the separation, dropping to $\sim$3.4\,mag for 
wide (50--1000\,mas) binaries. 
To confirm the limit, we performed a Bayesian model comparison by 
calculating the odds ratio between both models. Equal prior probability of each model was assumed 
(hence simplifying the problem to the calculation of the Bayes factor). 
The marginal likelihood (evidence) for each model 
was calculated using Nested Sampling \citep{skilling04}. We assumed a Gaussian 
prior probability for the primary diameter, and a uniform prior for the separation and contrast 
of the binary model. Due to computational reasons the calculation was performed for 150 
simulations and restricted to a smaller range of close-in (1--10\,mas) companions. 
The computed Bayes factors consistently favored the binary model over the single star model for 
contrasts $<$\,3.5\,mag. As expected, this value is more conservative than the limit inferred 
from the likelihood ratio, since the Bayesian model comparison penalizes the binary model for 
its added complexity. Based on these results, we conclude that our data would have revealed a stellar companion
at separations $\sim$\,1--1000\,mas ($\sim$\,0.1--113\,AU) and contrasts $<$\,3.5 magnitudes.

We note that the above simulations assume that our measurement uncertainties are well 
characterized. Previous PAVO observations of the same star over multiple nights 
showed that night-to-night variations in $V^2$ are at the 2--3\% level 
(Huber et al. 2012, in preparation), while the $\sim$\,5\,\% 
uncertainties in the calibrator sizes 
(see Table 1) translate to a 1\,\% uncertainy in $V^2$ \citep{vanbelle05}. Both these 
contributions are considerably smaller than our measurement uncertainties 
(estimated from the scatter of individual data frames integrated over each scan), which are on 
average 13\%. Combined with the low reduced $\chi^2$ of our fit, we therefore argue that our 
measurement uncertainties are a good estimate of the total uncertainty in our data.

The results presented here complement and extend the constraints set by \citet{howell12}. 
For the case of unbound companions, the PAVO limit of $<$\,3.5\,mag covers only a small 
range of the possible false-positive scenarios, which extend down to $<$\,7\,mag. Additionally,
\citet{howell12} demonstrated that the probability of a 
chance alignment of a star able to reproduce the transit shape at such close separations 
is very small. Hence, the added 
information from PAVO for unbound companions is negligible.
For the case of bound companions, Figure 3 illustrates the constraints from PAVO and speckle 
interferometry on the mass and separation of a possible secondary. 
Magnitude limits were converted into mass limits by 
interpolating a 3\,Gyr solar-metallicity isochrone by \citet{baraffe}, which roughly 
matches the age of HD\,179070 as determined by the asteroseismic analysis in \citet{howell12}. 
Additionally, we plot the 3-$\sigma$ detection limit from radial velocity follow-up 
for different inclinations, assuming a circular orbit for the hypothetical companion and a simple 
linear velocity change with time.
The diagram shows that PAVO rules out face-on orbits  
for bound companions at separations $\lesssim5.6$\,AU, which could previously not be 
excluded, and confirms the constraints set by complementary follow-up methods. 
Since a planet transiting a secondary with a mass lower than
0.8\,\msun\ can be excluded from the Kepler transit shape \citep{howell12}, the combined 
constraints virtually 
rule out any possible false-positive scenario involving a gravitationally bound companion around 
HD\,179070.

\begin{figure}
\begin{center}
\resizebox{\hsize}{!}{\includegraphics{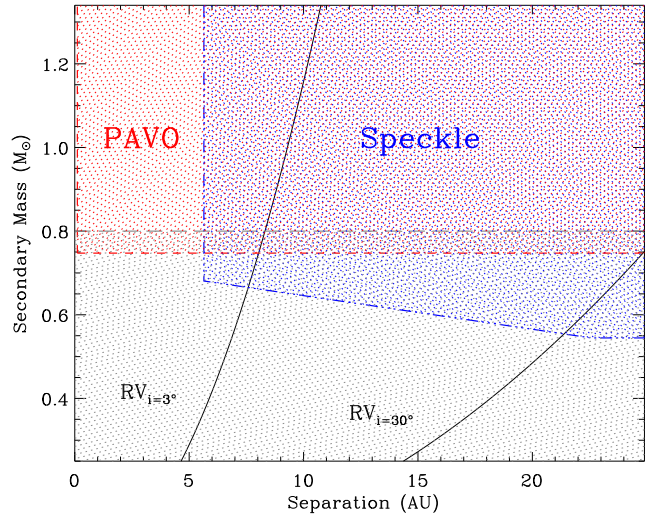}}
\caption{Constraints on the mass and separation of a secondary 
companion to HD\,179070 from PAVO (red dashed line) and 
Speckle interferometry (blue dashed-triple-dotted line). Magnitude limits have been 
converted into secondary masses by 
interpolating a 3\,Gyr solar-metallicity isochrone by \citet{baraffe}.
Black solid lines show the limit from 
RV follow-up for different inclinations, with any companion to the left of the lines 
being detectable with an average radial velocity signal $>16.8$\,m/s ($>\,3\,\sigma$) over a 
timespan of 85 days. Note that any false-positive
detection for secondaries with masses $\lesssim$\,0.8\,\msun\ (marked by a long-dashed line) 
can be ruled out from the observed Kepler transit shape \citep[see][]{howell12}.}
\label{fig3}
\end{center}
\end{figure}

\section{Conclusions}
We have presented high-angular-resolution observations of the exoplanet host star HD\,179070 
(Kepler-21) using the PAVO beam combiner at the CHARA Array. Our data clearly 
rule out stellar companions at separations between $\sim$\,1--1000\,mas ($\sim$\,0.1--113\,AU)
with contrasts of $<$\,3.5 magnitudes.
This complements and extends the validation efforts by \citet{howell12}, and supports the
conclusion that the 
detected transit is due to a 1.6\,$R_{\earth}$ extrasolar planet in an orbit 
around HD\,179070.

The results shown here demonstrate the potential of PAVO/CHARA to 
validate transiting exoplanet candidates, and complement the existing efforts using long-baseline 
interferometry to characterize exoplanet host stars 
\citep[see, e.g.,][]{baines09,vanbelle09,vonbraun11}. 
Using a recent compilation of detected exoplanets in the NASA Exoplanet 
Archive\footnote{http://exoplanetarchive.ipac.caltech.edu/index.html}, we 
estimate about half a dozen host stars with transiting exoplanets 
to be accessible to observations with PAVO/CHARA. 
Furthermore, there will be a considerable overlap with the target sample of the 
planned Transiting Exoplanet Survey Satellite \citep[TESS,][]{ricker09}, which is aimed at finding 
planets around nearby ($V<12$) stars. While the contribution of PAVO 
to the validation effort of Kepler-21b is relatively modest, it can be expected that long-baseline 
interferometry will play a significant role in validating transiting extrasolar planets, 
in particular for future missions aimed at bright stars and for cases where precise 
radial-velocity follow-up may not be available.

\section*{Acknowledgments}
DH, TRB and VM acknowledge support from the Access to Major Research Facilities Program, 
administered by the Australian Nuclear Science and Technology Organisation (ANSTO). 
DH is supported by an appointment to the NASA Postdoctoral Program at Ames Research Center, 
administered by Oak Ridge Associated Universities through a contract with NASA.
The CHARA Array is funded by the National Science Foundation through NSF grant AST-0606958, by 
Georgia State University through the College of Arts and Sciences, and by the W.M. Keck Foundation. 
This publication makes use of data products from the Two Micron All Sky Survey, which is a joint 
project of the University of Massachusetts and the Infrared Processing and Analysis Center/California 
Institute of Technology, funded by the National Aeronautics and Space Administration and the National 
Science Foundation. This research has made use of the NASA Exoplanet Archive, which is operated by 
the California Institute of Technology, under contract with the National Aeronautics and Space 
Administration under the Exoplanet Exploration Program.

\bibliographystyle{mn2e}
\bibliography{/Users/daniel/science/codes/latex/references}

\end{document}